\documentclass[12pt,twoside,dvips]{article}
\usepackage{amssymb}
\usepackage{amsfonts}
\usepackage{amsmath}
\usepackage{pifont}
\usepackage[spanish,english]{babel}
\usepackage{graphicx}
\usepackage{verbatim}

\begin{document}

\title{Quark mass hierarchy in 3-3-1 models}
\author{C. Alvarado, R. Mart\'{\i}nez$\thanks{%
e-mail: remartinezm@unal.edu.co}$, F. Ochoa$\thanks{%
e-mail: faochoap@unal.edu.co}$, \and Departamento de F\'{\i}sica, Universidad
Nacional de Colombia, \\ Ciudad Universitaria, Bogot\'{a} D.C.}
%EndAName
 
\maketitle

\begin{abstract}
We study the mass spectrum of the quark sector in an special type I-like model with gauge symmetry $SU(3)_c \otimes SU(3)_L \otimes U(1)_X$. By considering couplings with scalar triplets at large ($\sim TeV$) and small ($\sim GeV$) scales, we obtain specific zero-texture mass matrices for the quarks which predict three massless quarks ($u,d,s$) and three massive quarks ($c,b,t$)  at the electroweak scale ($\sim$ GeV). Taking into account mixing couplings with three heavy quarks at large scales predicted by the model, the three massless quarks obtain masses at small order that depends on the inverse of the large scale. Thus, masses of the form $m_u \lesssim m_d < m_s \sim MeV$ and $ m_{c,b,t} \sim GeV$ can be obtained naturally from the gauge structure of the model.  
\end{abstract}

\section{Introduction}

Although the Standard Model (SM) \cite{SM} is the simplest model that succesfully explain most of the phenomena and experimental observations in particle physics, it contains unanswered fundamental questions which many theorists associate to an underlying theory beyond the SM. In particular, the observed fermion mass hierarchies, their mixing and the three family structure are not explained in the SM. From the phenomenological point of view, it is possible to describe some features of the mass hierarchy by assuming zero-texture Yukawa matrices \cite{textures}. Models with spontaneously broken flavor symmetries may also produce hierarchical mass structures. These horizontal symmetries can be continuos and Abelian, as the original Froggatt-Nielsen model \cite{froggatt}, or non-Abelian as for example SU(3) and SO(3) family models \cite{non-abelian}. Models with discrete symmetries may also predict mass hierarchies for leptons \cite{discrete-lepton} and quarks \cite{discrete-quark}. Other models with horizontal symmetries have been proposed in the literature \cite{horizontal}.     

On the other hand, the origin of the family structure of the fermions can be addressed in family dependent models where a symmetry distinguish fermions of different families. An interesting alternative that may provide a clue to this puzzle are the models with gauge symmetry $SU(3)_c \otimes SU(3)_L \otimes U(1)_X$, also called 3-3-1 models, which introduce a family non-universal $U(1)_X$ symmetry \cite{331-pisano, 331-frampton, 331-long, M-O}. These models have a number of phenomenological advantages. First of all, from the cancellation of chiral anomalies \cite{anomalias} and
asymptotic freedom in QCD, the 3-3-1 models can explain why there are three fermion
families. Secondly, since the third family is treated under a different representation, the
large mass difference between the heaviest quark family and the two lighter ones may be
understood \cite{third-family}.  Also, these models contain a natural Peccei-Quinn symmetry, necessary to solve the strong-CP problem \cite{PC}.

In particular, the 3-3-1 models introduce three $SU(3)_L$ scalar triplets:  one heavy triplet field with a Vacuum Expectation Value (VEV) at high energy scale $\langle \chi \rangle = \nu _{\chi}$, which produces the breaking of the symmetry $SU(3)_L \otimes U(1)_X$ into the SM electroweak group $SU(2)_L \otimes U(1)_Y$, and two lighter triplets with VEVs at the electroweak scale  $\langle \rho \rangle = \upsilon _{\rho}$ and $\langle \eta \rangle = \upsilon _{\eta}$, which induce the electroweak breakdown. Thus, the model may provide masses to all fermions and gauge bosons at tree level. On the other hand, the 3-3-1 model possess an specialized Two Higgs Doublet Model type III (2HDM-III) in the low energy limit, where both electroweak triplets $\rho$ and $\eta$ are decomposed into two hypercharge-one $SU(2)_L$ doublets plus charged and neutral singlets. %Thus, like the 2HDM, the 3-3-1 model can be classified according to the way in which the Yukawa mixing matrices rotate, generating four types of models (type $a, b, c$ and $d$) \cite{rodolfo} which can be separated into terms free from Flavor Changing Neutral Currents (FCNC) plus FCNC couplings. In the first case ($a$), the mass matrices and Yukawa couplings of the up- and down-type quarks with the triplet $\rho$ ($h_{\rho}^{U,D}$) are simultaneously diagonalized. In type $b$, the Yukawa constants $h_{\rho}^{U}$ and $h_{\eta}^{D}$ diagonalize with the mass matrices. The type $c$ and $d$ are the same as $a$ and $b$ but changing $\rho \leftrightarrow \eta$. 
Thus, like the 2HDM-III, the 3-3-1 model can predict huge flavor changing neutral currents (FCNC) and CP-violating effects, which are severely suppressed by experimental data at electroweak scales. One way to remove these effects, is by imposing discrete symmetries, obtaining two types of 3-3-1 models (type I and II models), which exhibits the same Yukawa interactions as the 2HDM type I and II at low energy. In the first case, one Higgs electroweak triplet (for example, $\rho$) provide masses to the phenomenological up- and down-type quarks, simultaneously. In the type II, one Higgs triplet ($\rho$) gives masses to the up-type quarks and the other triplet ($\eta$) to the down-type quarks.  %In this paper we show that in the 331-I ...
In this paper we obtain in the framework of the $I$-type model specific mass matrix structures from the gauge symmetry, where only one of the down-type quarks acquire mass (which could be associated to the phenomenological bottom quark), and two are massless ($d$- and $s$-quarks), while two of the up-type quarks acquire masses ($c$- and $t$-quarks) and one is massless ($u$-quark). We also show by the method of recursive expansion \cite{grimus} that if mixing couplings with the heavy quark sector of the 3-3-1 model is considered, then the massless quarks indeed may obtain masses at small order that depends on the inverse of the heavy scale (represented by three heavy quarks) without introducing neither effective operators nor one-loop corrections \cite{effective}, \cite{effective-331}. Thus, at first glance we can obtain masses with the structures $m_u \lesssim m_d < m_s \sim MeV$ and $m_b \sim m_c \sim m_t \sim GeV$.    

This paper is organized as follows. In Section 2 we briefly describe some theoretical aspects of the 3-3-1 model and its particle content, in particular in the fermionic and scalar sector in order to obtain the mass spectrum. Section 3 is devoted to obtain the mass matrices in the low energy limit. In section 4 we consider the method of recursive expansion to diagonalize the mass matrices taking into account mixing couplings between light and heavy fermions. Finally in Sec. 5, we summarize and discuss our results.

\section{The Yukawa couplings of the 3-3-1 model}

We consider a 3-3-1 model where the electric charge is defined by:

\begin{eqnarray}
Q=T_3-\frac{1}{\sqrt{3}} T_8+X,
\end{eqnarray}
with $T_3=\frac{1}{2}Diag(1,-1,0)$ and $T_8=(\frac{1}{2\sqrt{3}})Diag(1,1,-2)$. In order to avoid chiral anomalies, the model introduces in the fermionic sector the following $(SU(3)_c, SU(3)_L,U(1)_X)$ left-handed representations: one $(3,3,1/3)$ quark triplet, two $(3,3^*,0)$ quark triplets and three $(1,3,-1/3)$ lepton triplets. For the right-handed sector, we introduce the following singlets in order to obtain Dirac-type charged fermions: three $(3^*,1,-1/3)$ down-type quarks, three $(3^*,1,2/3)$ up-type quarks, three $(1,1,-1)$ electron-type leptons. In addition we introduce three $(3^*,1,Q_{J_{1,2},T_1})$ and three $(1,1,0)$ right-handed singlets associated to the new non-SM quarks and neutral Majorana leptons, respectively.  In summary, we have the following representations free from chiral anomalies:

\begin{eqnarray}
Q^{1}_L
&=&
\begin{pmatrix}
U^{1} \\
D^{1} \\
T^{1} \\
\end{pmatrix}_L:(3,3,1/3) , 
\left\{
\begin{array}{c}
U^{1}_R :(3^*,1,2/3) \\ 
D^{1}_R :(3^*,1,-1/3) \\ 
T^{1}_R :(3^*,1,2/3) \\
\end{array}
\right.  \nonumber \\
Q^{2,3}_L
&=&
\begin{pmatrix}
D^{2,3} \\
U^{2,3} \\
J^{2,3} \\
\end{pmatrix}_L:(3,3^*,0), \left\{
\begin{array}{c}
D^{2,3}_R :(3^*,-1/3) \\
U^{2,3}_R :(3^*,1,2/3) \\
J^{2,3}_R :(3^*,1,-1/3) \\
\end{array}
\right. \nonumber  \\
L^{1,2,3}_L
&=&
\begin{pmatrix}
\nu ^{1,2,3} \\
e^{1,2,3} \\
(\nu^{1,2,3})^c \\
\end{pmatrix}_L :(1,3,-1/3) ,\left\{
\begin{array}{c}
e^{1,2,3}_R :(1,1,-1) \\
N_R^{1,2,3}:(1,1,0) \\
\end{array}
\right.
\label{fermion_spectrum}
\end{eqnarray}
where $U^{i}_L$ and $D^{i}_L$ for $i=1,2,3$ are three up- and down-type quark components in the flavor basis,  while $\nu ^{i}_L$ and $e^{i}_L$ are the neutral and charged lepton families. The right-handed sector transform as singlets under $SU(3)_L$ with $U(1)_X$ quantum numbers equal to the electric charges. In addition, we see that the model introduces heavy fermions with the following properties: a single flavor quark  $T^{1}$ with electric charge $2/3$, two flavor quarks $J^{2,3}$ with charge $-1/3$, three neutral Majorana leptons $(\nu^{1,2,3})^c_L$ and three right-handed Majorana leptons $N^{1,2,3}_R$ (recently, a discussion about neutrino masses via double and inverse see-saw mechanism was perform in ref. \cite{catano}).  On the other hand, the scalar sector introduces one triplet field with VEV $\langle \chi \rangle_0=\upsilon_{\chi}$, which provides the masses to the new heavy fermions, and two triplets with VEVs $\langle \rho \rangle_0=\upsilon_{\rho}$ and $\langle \eta \rangle_0=\upsilon_{\eta}$, which give masses to the SM-fermions at the electroweak scale. The $(SU(3)_L,U(1)_X)$ group structure of the scalar fields are:

\begin{eqnarray}
\chi&=&
\begin{pmatrix}
\chi_1^{0}\\
\chi_2^{-} \\
\frac{1}{\sqrt{2}}(\upsilon_{\chi} + \xi_{\chi} \pm i \zeta_{\chi} ) \\
\end{pmatrix}: (3,-1/3) \notag \\
\rho&=&
\begin{pmatrix}
\rho_1^{+} \\
\frac{1}{\sqrt{2}}(\upsilon_{\rho} + \xi_{\rho} \pm i \zeta_{\rho} ) \\
\rho _3^{+} \\
\end{pmatrix} : (3,2/3) \notag \\
\eta &=&
\begin{pmatrix}
\frac{1}{\sqrt{2}}(\upsilon_{\eta} + \xi_{\eta} \pm i \zeta_{\eta} ) \\
\eta _2^{-} \\
\eta _3^{0}
\end{pmatrix}:(3,-1/3).
\label{331-scalar} 
\end{eqnarray}

With the above spectrum, we obtain the following $SU(3)_L \otimes U(1)_X$ renormalizable Yukawa Lagrangian for the quark sector:

\begin{eqnarray}
-\mathcal{L}_Y &=& \overline{Q_L^{1}}\left(\eta h^{U}_{\eta 1j}+\chi h^{U}_{\chi 1j}\right)U_R^{j}+
\overline{Q_L^{1}}\rho h^{D}_{\rho 1j}D_R^{j}\notag \\
&+&\overline{Q_L^{1}} \rho h^{J}_{\rho 1m} J^{m}_R+\overline{Q_L^{1}}\left(\eta h^{T}_{\eta 11}+\chi h^{T}_{\chi 11}\right)T_R^{1} \notag \\
&+&\overline{Q_L^{n}}\rho ^* h^{U}_{\rho nj}U_R^{j}+\overline{Q_L^{n}}\left(\eta ^* h^{D}_{\eta nj}+\chi ^* h^{D}_{\chi nj}\right)D_R^{j} \notag \\
&+&\overline{Q_L^{n}}\left( \eta ^* h^{J}_{\eta nm}+\chi ^* h^{J}_{\chi nm}\right) J^{m}_R+\overline{Q_L^{n}} \rho^* h^{T}_{\rho n1}T_R^{1}
 + h.c,
 \label{331-yukawa}
\end{eqnarray}
where $n=2,3$ is the index that label the second and third quark triplet shown in Eq. (\ref{fermion_spectrum}), and $h^{f}_{\phi ij}$ are the ${i,j}$ components of non-diagonal matrices in the flavor space associated with each scalar triplet $\phi : \eta , \rho, \chi$. In order to avoid FCNC terms at tree level, we demand the following discrete symmetry:

\begin{eqnarray}
\eta &\rightarrow& -\eta, \hspace{0.3cm} \rho \rightarrow \rho, \notag \\
D_R &\rightarrow& D_R, \hspace{0.3cm} U_R \rightarrow U_R, \notag \\
T_R &\rightarrow& T_R, \hspace{0.3cm} J_R \rightarrow J_R.
\label{discrete}
\end{eqnarray}
Thus, the couplings of the quarks with the triplet $\eta$ are removed from the Lagrangian in (\ref{331-yukawa}), which at low energy is equivalent to 2HDM type I. After the symmetry breaking of the 3-3-1 gauge group, we obtain from the VEVs in (\ref{331-scalar}) and taking into account the Yukawa Lagrangian in (\ref{331-yukawa}) with the discrete symmetries in (\ref{discrete}) the following structure for the mass terms:

\begin{eqnarray}
-\langle \mathcal{L}_Y \rangle =
\left(\overline{U_L},\overline{T_L}\right)M_{UT}
\begin{pmatrix}
U_R \\
T_R
\end{pmatrix}+\left(\overline{D_L},\overline{J_L}\right)M_{DJ}
\begin{pmatrix}
D_R \\
J_R
\end{pmatrix}
+h.c,
\label{mass-yukawa}
\end{eqnarray}
where $U_{L,R}=(U^{1},U^{2},U^{3})_{L,R}$ are the left- and right-handed up-type quark flavor vectors,  $D_{L,R}=(D^{1},D^{2},D^{3})_{L,R}$ the corresponding down-type quark vectors, $J_{L,R}=(J^{2},J^{3})_{L,R}$ are two-dimensional vectors associated with the heavy quarks with electric charge $-1/3$ in (\ref{fermion_spectrum}) and $T_{L,R}$ is the single component of the heavy quark with charge $2/3$. The matrices $M_{UT}$ and $M_{DJ}$ have the following structures in the basis $(U,T)$ and $(D,J)$, respectively:   

\begin{eqnarray}
M_{UT}=\begin{pmatrix}
M_U && k \\
K && M_T \\
\end{pmatrix}, \hspace{0.3cm}
M_{DJ}=\begin{pmatrix}
M_D && s \\
S && M_J \\
\end{pmatrix},
\label{mixing-mass}
\end{eqnarray} 
where $M_{U}$, $k$, $K$ and $M_T$ are $3 \times 3$, $3 \times 1$, $1 \times 3$ and $1 \times 1$ matrices, respectively, while $M_{D}$, $s$, $S$ and $M_J$ are $3 \times 3$, $3 \times 2$, $2 \times 3$ and $2 \times 2$ matrices, respectively. The Yukawa Lagrangian in (\ref{331-yukawa}) provides the following relations between the above mass matrices and the Yukawa couplings through the VEVs:

%\begin{eqnarray}
%M_U&=&\frac{1}{\sqrt{2}}\left(h_{\rho}^{U}\upsilon _{\rho}+h_{\eta}^{U} \upsilon _{\eta} \right)  ,\hspace{0.5cm}    M_T=\frac{1}{\sqrt{2}}h_{\chi }^{T}\upsilon _{\chi},\notag \\
%k&=&\frac{1}{\sqrt{2}}\left(h_{\rho}^{T}\upsilon _{\rho}+h_{\eta}^{T} \upsilon _{\eta} \right)  , \hspace{0.5cm} K=\frac{1}{\sqrt{2}}h_{\chi }^{U}\upsilon _{\chi}, 
%\label{UJ-mass}
%\end{eqnarray}
\begin{eqnarray}
M_U&=&\frac{1}{\sqrt{2}}h_{\rho}^{U}\upsilon _{\rho}   ,\hspace{0.5cm}    M_T=\frac{1}{\sqrt{2}}h_{\chi }^{T}\upsilon _{\chi},\notag \\
k&=&\frac{1}{\sqrt{2}}h_{\rho}^{T}\upsilon _{\rho}   , \hspace{0.5cm} K=\frac{1}{\sqrt{2}}h_{\chi }^{U}\upsilon _{\chi}, 
\label{UJ-mass}
\end{eqnarray}
for $M_{UT}$, and

%\begin{eqnarray}
%M_D&=&\frac{1}{\sqrt{2}}\left(h_{\rho}^{D}\upsilon _{\rho}+h_{\eta}^{D} \upsilon _{\eta} \right), \hspace{0.5cm}  M_J=\frac{1}{\sqrt{2}}h_{\chi }^{J}\upsilon _{\chi},\notag \\
%s&=&\frac{1}{\sqrt{2}}\left(h_{\rho}^{J}\upsilon _{\rho}+h_{\eta}^{J} \upsilon _{\eta} \right), \hspace{0.5cm}  S=\frac{1}{\sqrt{2}}h_{\chi }^{D}\upsilon _{\chi}, 
%\label{DT-mass}
%\end{eqnarray}

\begin{eqnarray}
M_D&=&\frac{1}{\sqrt{2}}h_{\rho}^{D}\upsilon _{\rho} , \hspace{0.5cm}  M_J=\frac{1}{\sqrt{2}}h_{\chi }^{J}\upsilon _{\chi},\notag \\
s&=&\frac{1}{\sqrt{2}}h_{\rho}^{J}\upsilon _{\rho} , \hspace{0.5cm}  S=\frac{1}{\sqrt{2}}h_{\chi }^{D}\upsilon _{\chi}, 
\label{DT-mass}
\end{eqnarray}
for $M_{DJ}$. We can see in the Yukawa Lagrangian in Eq. (\ref{331-yukawa}) that due to the non-universal form of the $U(1)_X$ values exhibited by the scalar and quark triplets in (\ref{fermion_spectrum}) and (\ref{331-scalar}), not all couplings between quarks and scalars are allowed by the gauge symmetry, which lead us to the following zero-texture Yukawa coupling constants:

\begin{eqnarray}
h_{\rho}^{U}=\begin{pmatrix}
0 && 0 && 0 \\
a && b && c \\
d && e && f \\
\end{pmatrix}, \hspace{0.3cm}
%h_{\eta}^{U}=\begin{pmatrix}
%a' && b' && c' \\
%0 && 0 && 0 \\
%0 && 0 && 0 \\
%\end{pmatrix},  \hspace{0.3cm}
h_{\chi}^{U}=
(a'', b'',c''),
\label{up-textures}
\end{eqnarray}
for the couplings with up-type quarks,

\begin{eqnarray}
h_{\rho}^{D}=\begin{pmatrix}
A && B && C \\
0 && 0 && 0 \\
0 && 0 && 0 \\
\end{pmatrix},  \hspace{0.3cm}
%h_{\eta}^{D}=\begin{pmatrix}
%0 && 0 && 0 \\
%A' && B' && C' \\
%D' && E' && F' \\
%\end{pmatrix},  \hspace{0.3cm}
h_{\chi}^{D}=\begin{pmatrix}
A'' && B'' && C'' \\
D'' && E'' && F'' \\
\end{pmatrix},
\label{down-textures}
\end{eqnarray}
for the couplings with down-type quarks,

\begin{eqnarray}
h_{\rho}^{J}=\begin{pmatrix}
w && x \\
0 && 0  \\
0 && 0  \\
\end{pmatrix},  \hspace{0.3cm}
%h_{\eta}^{J}=\begin{pmatrix}
%0 && 0  \\
%w' && x'  \\
%y' && z'  \\
%\end{pmatrix},  \hspace{0.3cm}
h_{\chi}^{J}=\begin{pmatrix}
w'' && x''  \\
y'' && z''  \\
\end{pmatrix},
\end{eqnarray}
for the couplings with the doublets $J$, and

\begin{eqnarray}
h_{\rho}^{T}=\begin{pmatrix}
0 \\
W  \\
X  \\
\end{pmatrix},  \hspace{0.3cm}
%h_{\eta}^{T}=\begin{pmatrix}
%W' \\
%0 \\
%0  \\
%\end{pmatrix},  \hspace{0.3cm}
h_{\chi}^{T}=
W'',
\label{T-textures}
\end{eqnarray}
for the couplings with the single $T$ quark. 
  
\section{The mass matrices in the low energy limit}

In the low energy limit ($\upsilon _{\chi} \gg \upsilon _{\eta,\rho}$), the quark mass eigenstates for the small and large scales can be obtained separately  by unitary transformations of the left- and right-handed weak eigenstates: $U'_{L,R}=V^{U\dagger}_{L,R}U_{L,R}$,  $D'_{L,R}=V^{D\dagger}_{L,R}D_{L,R}$, and $J'_{L,R}=V^{J\dagger}_{L,R}J_{L,R}$, while the singlet $T$-quark decouple from other components, obtaining $T'_{L,R}=T_{L,R}$. Thus, the matrices for $U$-,$T$-,$D$- and $J$-quarks in Eqs. (\ref{UJ-mass}) and (\ref{DT-mass}) are diagonalized by:

 \begin{eqnarray}
m_{U}&=&V_L^{U\dagger}M_UV_R^U=\frac{\upsilon_{\rho}}{\sqrt{2}}V_L^{U\dagger}h^{U}_{\rho}V_R^U,  \nonumber \\
m_{D}&=&V_L^{D\dagger}M_DV_R^D=\frac{\upsilon_{\rho}}{\sqrt{2}}V_L^{D\dagger}h^{D}_{\rho}V_R^D, \nonumber \\
m_{J}&=&V_L^{J\dagger}M_JV_R^J=\frac{\upsilon_{\chi}}{\sqrt{2}}V_L^{J\dagger}h^{J}_{\chi} V_R^J, \nonumber \\
m_{T}&=&\frac{\upsilon_{\chi}}{\sqrt{2}}h^{T}_{\chi} ,
\label{diag-mass}
\end{eqnarray}
Thus, the mass matrices for $U$- and $D$-type quarks depend only on the $h_{\rho}$ Yukawa matrices, which from (\ref{UJ-mass}), (\ref{DT-mass}), (\ref{up-textures}) and (\ref{down-textures}) become: %is equivalent to take $h^{U,D}_{\eta}=0$ in the above lagrangian.  %Thus, without considering the FCNC terms, the mass matrices for $U$- and $D$-type quarks depend only on the $h_{\rho}$ Yukawa matrices, which from (\ref{UJ-mass}), (\ref{DT-mass}), (\ref{up-textures}) and (\ref{down-textures}) become:
%Considering Type I models is equivalent to take $h^{U,D}_{\eta}=0$ in the above lagrangian.(\ref{UJ-mass}) and (\ref{DT-mass}) 

\begin{equation}
M_{U}=\frac{\upsilon _{\rho}}{\sqrt{2}}\begin{pmatrix}
0 && 0 && 0 \\
a && b && c \\
d && e && f \\
\end{pmatrix}, \hspace{0.5cm}
M_{D}=\frac{\upsilon _{\rho}}{\sqrt{2}}\begin{pmatrix}
A && B && C \\
0 && 0 && 0 \\
0 && 0 && 0 \\
\end{pmatrix},
\label{mass-type-I}
\end{equation}   
which diagonalize through the bi-unitary transformations $V^{U,D}_{L,R}$. Let us evaluate the eigenvalues of the square mass matrices:

\subsection{Up-sector}

From (\ref{mass-type-I}), we obtain the following structure:

\begin{eqnarray}
M_UM_U^{\dagger}=\begin{pmatrix}
0 && 0 && 0 \\
0 && \alpha && \beta \\
0 && \beta^* && \gamma \\
\end{pmatrix},
\label{up-mass square}
\end{eqnarray}
where $\alpha, \beta$ and $\gamma$ are of the order $\upsilon _{\rho}^2 \sim (246$ GeV$)^2$. The above matrix exhibits one zero eigenvalue and diagonalize with only $V^U_{L}$:

\begin{eqnarray}
V^{U\dagger}_LM_UM_U^{\dagger}V^{U}_L=m_U^{2}=diag(0,m_2^2,m_3^2).
\end{eqnarray}
Thus, if we identify the zero-mass component with the phenomenological $u$-quark, and the other two with the $c$- and $t$-quarks, we obtain:

\begin{eqnarray}
m_u^2&=&0 \nonumber \\
m_{c,t}^2&=&m_{2,3}^2 \sim GeV^2.
\label{up-hierarchy}
\end{eqnarray}  

\subsection{Down-sector}

From (\ref{mass-type-I}), we obtain the matrix:

\begin{eqnarray}
M_DM_D^{\dagger}=\begin{pmatrix}
\Sigma && 0 && 0 \\
0 && 0 && 0 \\
0 && 0 && 0 \\
\end{pmatrix},
\label{down-mass square}
\end{eqnarray}
that exhibits two massless quarks which can be associated with the $d$- and $s$-quarks, while $\Sigma$ with the $b$-quark:

\begin{eqnarray}
m_{d,s}^2&=&0 \nonumber \\
m_{b}^2&=&\Sigma \sim GeV^2
\label{down-hierarchy}
\end{eqnarray} 
Thus, from the gauge symmetry, we may obtain zero-texture mass matrices for the quark sector in the low energy limit, obtaining three massless quarks (light quarks) and three massive quarks at the electroweak scale ($\sim$ GeV). The massless quarks are indeed massive particles if we consider small couplings with the extra heavy 3-3-1 quarks, as we show below.

\section{The mass matrices with mixing couplings}
 
In this case we consider the complete mass matrices from Eq. (\ref{mixing-mass}), which have the following general structure:

\begin{eqnarray}
M=\begin{pmatrix}
M_{light} && f_{light}  \\
G_{heavy} && \Lambda _{heavy}  \\
\end{pmatrix},
\label{mixing-matrix}
\end{eqnarray}
where $M_{light} \sim f_{light} \sim \upsilon_{\rho} \sim 246$ GeV, while $G_{heavy}\sim \Lambda_{heavy}\sim \upsilon _{\chi} \gg 246$ GeV. The above mass matrix can be diagonalized by a bi-unitary transformation: $\widetilde{m}=\mathcal{O}_L^{\dagger}M\mathcal{O}_R$. This transformation can be separated into two rotations: first, we can rotate through the bi-unitary transformations $V_{L,R} = V_{L,R}^{U,D}$ and $P_{L,R}=V_{L,R}^{J}$ defined by Eq. (\ref{diag-mass}) in the low energy limit. Second, since the first rotation does not lead to a completely diagonal matrix due to the mixing terms $f$ and $G$, we must perform another rotation through unitary matrices $B_{L,R}$. Thus, we separate the original rotation into:
 
\begin{eqnarray}
\mathcal{O}_{L,R}&=&\mathcal{U}_{L,R}W_{L,R}=\begin{pmatrix}
V_{L,R} && 0 \\
0 && P_{L,R}\\
\end{pmatrix}\begin{pmatrix}
1 && B_{L,R} \\
-B_{L,R}^{\dagger} && 1\\
\end{pmatrix},
\label{double-rotation}
\end{eqnarray}
where we require that:

\begin{eqnarray}
\mathcal{O}_{L}^{\dagger}M\mathcal{O}_R=W_{L}^{\dagger}\mathcal{U}_{L}^{\dagger}M\mathcal{U}_RW_R=\widetilde{m}=\begin{pmatrix}
\widetilde{m} _{light} && 0\\
0 && \widetilde{m} _{heavy}\\
\end{pmatrix}.
\label{diagonal-total}
\end{eqnarray}
 
After the first rotation (through $\mathcal{U}_{L,R}$) of (\ref{mixing-matrix}), we obtain mixing matrices of the form:
 
\begin{eqnarray}
\mathcal{U}_{L}^{\dagger}M\mathcal{U}_R=m=\begin{pmatrix}
m_{l} && \widetilde{f} \\
\widetilde{G} && M_{H} \\
\end{pmatrix},
\label{first-mixing-mass}
\end{eqnarray} 
where $m_{l}=V_L^{\dagger}M_{light}V_R$ and $M_{H}=P_L^{\dagger}\Lambda_{heavy}P_R$ are diagonal blocks, while $\widetilde{f}=V_L^{\dagger}f_{light}P_R$ and $\widetilde{G}=P_L^{\dagger}G_{heavy}V_R$ are non-diagonal mixing blocks. Then, we must find the matrices $B_{L,R}$ in order to obtain the complete diagonalization of the mixing matrices in (\ref{first-mixing-mass}).  We can achieve this by constructing the following squared mass matrices from (\ref{first-mixing-mass}):

\begin{eqnarray}
m_1^2=mm^{\dagger}&=&\begin{pmatrix}
m_{l}m_{l}^{\dagger}+\widetilde{f}\widetilde{f}^{\dagger} && m_{l}\widetilde{G}^{\dagger}+\widetilde{f}M_{H}^{\dagger}  \\
\widetilde{G}m_{l}^{\dagger}+M_{H}\widetilde{f}^{\dagger}  && M_{H}M_{H}^{\dagger}+\widetilde{G}\widetilde{G}^{\dagger} \\
\end{pmatrix} =\begin{pmatrix}
a _l && x_{m} \\
x_{m}^{\dagger} &&  y_H \\
\end{pmatrix}  \nonumber \\
m_2^2=m^{\dagger}m&=&\begin{pmatrix}
m_{l}^{\dagger}m_{l}+ \widetilde{G}^{\dagger}\widetilde{G} && m_{l}^{\dagger}\widetilde{f}+\widetilde{G}^{\dagger}M_{H} \\
\widetilde{f}^{\dagger}m_{l}+M_{H}^{\dagger}\widetilde{G}  && M_{H}^{\dagger}M_{H}+\widetilde{f}^{\dagger}\widetilde{f} \\
\end{pmatrix}=\begin{pmatrix}
A_H && X _{H} \\
X_{H}^{\dagger} && Y _H \\
\end{pmatrix} ,
\label{squared-matrix}
\end{eqnarray}
where $a_l \sim \upsilon _{\rho}^2$ (electroweak scale), $x_m \sim \upsilon _{\rho}\upsilon _{\chi}$ (intermediate scale) and $y_H \sim A_H \sim X_H \sim {Y_H} \sim \upsilon _{\chi}^2$ (heavy scale). The above squared matrices are diagonalized through $W_{L,R}$ defined in (\ref{double-rotation}):

\begin{eqnarray}
W_{L,R}^{\dagger}m_{1,2}^2 W_{L,R}=\begin{pmatrix}
\widetilde{m}_l^2 && 0 \\
 0 && \widetilde{M}_H^{2} \end{pmatrix}.
\label{second-rotation}
\end{eqnarray}
From the condition of the vanishing of the off-diagonal submatrices in Eq. (\ref{second-rotation}), we obtain:

\begin{eqnarray}
B_L(x_m^{\dagger})B_L-a_lB_L+B_Ly_H-x_m&=&0,
\label{second-L-rotation}
\end{eqnarray} 
\begin{eqnarray}
B_R(X_H^{\dagger})B_R-A_HB_R+B_RY_H-X_H&=&0.
\label{second-R-rotation}
\end{eqnarray} 
Since $a_l \ll x_m \ll y_H$, the equation (\ref{second-L-rotation}) may be  solved assuming that $B_L$ expands in powers of $1/y_H$ \cite{grimus}:

\begin{equation}
B_L=B_{L1}+B_{L2}+B_{L3}+.....
\end{equation}
where at order $B_{L1}$ Eq. (\ref{second-L-rotation}) approximates to $B_Ly_H-x_m=0$, obtaining:

\begin{equation}
B_{L}\approx x_my_H^{-1}=\left(m_{l}\widetilde{G}^{\dagger}+\widetilde{f}M_{H}^{\dagger}\right)\left(M_{H}M_{H}^{\dagger}+\widetilde{G}\widetilde{G}^{\dagger}\right)^{-1}.
\label{solution-L}
\end{equation}

Solving Eq. (\ref{second-R-rotation}) is less evident, since all coefficients are at the heavy scales. However, we may consider an scenario where the mixing terms in (\ref{mixing-matrix}) are small with respect to the diagonal components, which implies for the matrix $m_2^2$ in (\ref{squared-matrix}) the hierarchy $Y_H \gg X_H \gg A_H$. Thus, Eq.   (\ref{second-R-rotation}) may also be solved assuming that $B_R$ expands in powers of $1/Y_H$, where at first order Eq.  (\ref{second-R-rotation}) reads $B_RY_H-X_H=0$, obtaining:

\begin{equation}
B_{R}\approx X_HY_H^{-1}=\left(\widetilde{G}^{\dagger}M_{H}\right)\left(M_{H}^{\dagger}M_{H}\right)^{-1}.
\label{solution-R}
\end{equation}

Putting all together into the total rotation in Eq. (\ref{diagonal-total}), we finally find the light and heavy diagonal masses:

\begin{eqnarray}
\widetilde{m}_{light}&=&m_l-\widetilde{f}B_R^{\dagger}-B_L\widetilde{G}+B_LM_HB_R^{\dagger} \nonumber \\
\widetilde{m}_{heavy}&=&M_H+\widetilde{G}B_R+B_L^{\dagger}\widetilde{f}+B_L^{\dagger}m_lB_R.
\label{masses}
\end{eqnarray}
In particular, for the light sector we see that:

\begin{eqnarray}
m_l&=&V_L^{\dagger}M_{light}V_R, \nonumber \\
M_H&=&P_L^{\dagger}\Lambda_{heavy}P_R, \nonumber \\
B_L&\approx& \widetilde{f}/M_H, \nonumber \\
B_R^{\dagger}&\approx& \widetilde{G}/M_H,
%(Y_H^{-1})^{\dagger}&=&1/M_H^2, \nonumber \\
%y_H^{-1}&\approx &1/M_H^2, \nonumber \\
%x_m&\approx& \widetilde{f}M_H^{\dagger} 
\end{eqnarray}
obtaining for the light fermions in (\ref{masses}):

\begin{equation}
\widetilde{m}_{light} \approx V_L^{\dagger}M_{light}V_R-\frac{\widetilde{f}\widetilde{G}}{M_H}.
\end{equation}
If we apply the above solution into the mass matrices in (\ref{mixing-mass}) we will obtain for the light sector that:

\begin{eqnarray}
\widetilde{m}_{U} &\approx& V_L^{U\dagger}M_{U}V_R^U-\frac{\widetilde{k}\widetilde{K}}{M_T}, \\
\widetilde{m}_{D} &\approx& V_L^{D\dagger}M_{D}V_R^D-\frac{\widetilde{s}\widetilde{S}}{M_J},
\end{eqnarray}
where the first terms correspond to the diagonal masses in the low energy limit given by (\ref{up-hierarchy}) and (\ref{down-hierarchy}), plus small corrections that arise from the mixing terms $k,s,K,S$ and the inverse of the heavy masses of the T- and J-quarks. Thus, if we make the same identification as in  (\ref{up-hierarchy})  and (\ref{down-hierarchy}), we obtain for the up sector:

\begin{eqnarray}
\left|\widetilde{m}_{u}\right| &\approx& \frac{\left|\widetilde{k}\widetilde{K}\right|_{11}}{M_T} \sim MeV \nonumber \\ 
\left|\widetilde{m}_{c,t}\right| &\approx& \left|V_L^{U\dagger}M_{U}V_R^U\right|_{22,33} \sim GeV,
\label{hierarchy-complete-up}
\end{eqnarray}
while for the down sector:

\begin{eqnarray}
\left|\widetilde{m}_{d,s}\right| &\approx& \frac{\left|\widetilde{s}\widetilde{S}\right|_{22,33}}{M_{J_{2,3}}} \sim MeV \nonumber \\
\left|\widetilde{m}_{b}\right| &\approx &\left|V_L^{D\dagger}M_{D}V_R^D\right|_{11} \sim GeV.
\label{hierarchy-complete-down}
\end{eqnarray}
Indeed, we see in (\ref{hierarchy-complete-down}) that $\widetilde{m}_{d,s}$ depends on the inverse of the two masses of the J-quarks, while in (\ref{hierarchy-complete-up}), $\widetilde{m}_u$ is inverse in $M_T$. Thus, if we require that the heavy quarks obey $M_T \gtrsim M_{J_{2}} > M_{J_{3}}$ we obtain the following forms:

\begin{eqnarray}
\widetilde{m}_u \lesssim \widetilde{m}_d < \widetilde{m}_s &\sim& MeV \nonumber \\
 \widetilde{m}_{b,c,t} &\sim& GeV.
\end{eqnarray}

\section{Conclusions}

The 3-3-1 model exhibits an abelian non-universal $U(1)_X$ symmetry in the quark sector, which lead us that not all Yukawa couplings are allowed by the symmetry. Indeed, the family-dependence feature shown by the quark multiplets in (\ref{fermion_spectrum}) arises from the condition of cancellation of the chiral anomalies in order to obtain a realistic renormalizable spectrum beyond the tree-level.  Thus, from the gauge structure of the model, we obtain zero-texture Yukawa coupling constants $h_{\rho,\chi}$ as shown by equations (\ref{up-textures})-(\ref{T-textures}). These structures may generate quark mass hierarchies if we consider an special basis through appropriate discrete symmetries that suppress the FCNC couplings, analogous to the 2HDM type I. In this case, one Higgs triplet ($\rho$) provides masses to the up- and down-type quarks simultaneously, obtaining zero-texture mass matrices through the VEV $\upsilon _{\rho}$, as shown by (\ref{mass-type-I}). The above matrices exhibits one massless up-type quark ($u$-quark) and two massless down-type quarks ($d$- and $s$-quarks), while three quarks ($c$-, $b$- and $t$-quarks) have masses at the scale $\upsilon _{\rho} \sim$ GeV.  %and II:
\vspace{0.5cm}

%We see that the model that best describe the phenomenological mass hierarchy is the type-I model.
On the other hand, we may generate small ($\sim$ MeV) mass components to the above massless quarks without introducing neither effective operators nor one-loop corrections. If we consider the complete allowed Yukawa couplings, including small mixing couplings with the heavy $T$-, $J_1$- and $J_2$-quarks (which according to (\ref{diag-mass}) have masses at large scale $\nu _{\chi} \sim$ TeV), the mixing mass matrices in (\ref{mixing-mass}) can be diagonalized into light and heavy masses. In particular, by the method of recursive expansion, it is possible to decouple both scales at first order, obtaining see-saw type masses, where the massless quarks acquire masses at scales inverse in the heavy mass quarks: $ |\widetilde{m}_u| \sim | \widetilde{k} \widetilde{K}|/M_T$, $ |\widetilde{m}_{d,s}| \sim | \widetilde{s} \widetilde{S}|/M_{J_{2,3}}$. If we consider heavy non-degenerated spectrum, in particular that $M_T \gtrsim M_{J_2} > M_{J_3}$ we may understand the observed hierarchy $m_u \lesssim m_d < m_s$ exhibited by the phenomenological light quark sector.

This work was supported by Colciencias.

\end{document}